\documentclass[twocolumn]{article}
\usepackage{xcolor}
\usepackage{braket}
\usepackage{comment}
\usepackage{amsfonts, amssymb, amsmath, amsthm, mathrsfs, chet,cite}
\usepackage{setspace}
\usepackage{titlesec}
\usepackage{balance}

\makeatletter
\let\c@lofdepth\relax
\let\c@lotdepth\relax
\makeatother
\usepackage{subfigure}
\usepackage{graphicx}
\usepackage{caption}
\usepackage{comment}
\usepackage{float}
\usepackage{afterpage}

\begin{document}
\title{Tensor Network Based Efficient Quantum Data Loading of Images}
\author{Jason Iaconis, Sonika Johri\footnote{Now at Coherent Computing Inc.}}
\affiliation{ IonQ Inc., 4505 Campus Drive, College Park, MD 20740, USA}
\abstract{Image-based data is a popular arena for testing quantum machine learning algorithms. A crucial factor in realizing quantum advantage for these applications is the ability to efficiently represent images as quantum states.  Here we present a novel method for creating quantum states that approximately encode images as amplitudes, based on recently proposed techniques that convert matrix product states to quantum circuits. The numbers of gates and qubits in our method scale logarithmically in the number of pixels given a desired accuracy, which make it suitable for near term quantum computers. Finally, we experimentally demonstrate our technique on 8 qubits of a trapped ion quantum computer for complex images of road scenes, making this the first large instance of full amplitude encoding of an image in a quantum state.}

\maketitle

Quantum machine learning is emerging as a promising avenue for the application of near-term quantum computers. Recent work has shown that quantum algorithms offer advantages in expressivity and efficiency for certain machine learning tasks \cite{Schuld_data_encoding, peters2022generalization, caro2022outofdistribution, Caro2022-ae, Lukin2022, bowles2023contextuality, PhysRevResearch.4.043092}, and thus have the potential to outperform their classical counterparts in specific domains. 

Images, as one of the most prevalent forms of data, have been extensively studied in classical machine learning. Quantum machine learning proposes new paradigms to accelerate image processing tasks.  Experimental demonstrations of image-based learning with quantum computers include the training of a quantum-enhanced generative adversarial network that generates images from the MNIST dataset  using 8 trapped-ion qubits \cite{Rudolph20}, a quantum nearest centroid algorithm on the MNIST dataset  on up to 8 trapped-ion qubits \cite{Johri20}, and classification of medical images 
on up to 6 superconducting qubits \cite{cherrat2022quantum}.

In any quantum algorithm that processes classical data, the step of representing that data as a quantum state is a crucial one. Efficient data loading is imperative for overall algorithmic performance, and in the context of quantum machine learning, it can affect whether and how much quantum advantage can be practically achieved \cite{Schuld_data_encoding}. Near-term quantum machine learning is usually formulated as a parametrized quantum circuit that is optimized according to a given learning task. In this framework, each data point is uploaded to a quantum state one at a time before the parametrized quantum circuit acts on it. In the case of images, each data point by itself can have a large amount of information, proportional to the number of pixels in the image. This poses a problem for near-term quantum computers because their gate fidelity is limited, with two-qubit gate fidelities typically an order of magnitude less than single-qubit gate fidelities. Overall, the number of gates in a quantum data loading circuits proposed so far scales with the size of the data. The number of two-qubit gates in particular is typically proportional to the `density' of the data storage, which can be defined as the ratio between the size of the classical data and the size of the Hilbert space.

Therefore, near-term quantum image processing algorithms often aim to represent data `sparsely', i.e. the number of qubits required scales linearly in the number of pixels. Examples of this are the unary amplitude encoding \cite{Johri20, cherrat2022quantum} in which the number of two-qubit gates and number of qubits is proportional to the data size, and product state encoding in which there may be no two-qubit gates involved in the encoding at all. However, since the number of qubits is also limited in near-term quantum computers, using these techniques means that images need to be compressed before loading using techniques like principal component analysis, variational autoencoders, or simple spatial averaging over image patches. In this process, one may lose information that is critical to the learning task, especially since none of these techniques are particularly sensitive to image-specific features like the presence of edges which may make it hard to do more complex image processing tasks such as object detection.

Ideally, therefore, there would exist an efficient method that can create a quantum state that `densely' stores the image data, i.e. the size of the Hilbert space is proportional to the number of pixels, and does not require many additional qubits during the state preparation procedure. In this context, the most recent proposal has been the QPIXL framework \cite{qpixl} in which the number of gates scales linearly in the number of pixels. The authors also propose a compression technique which involves setting small angles to 0 during the state preparation procedure and show that some images can be stored with high quality while significantly reducing the number of gates required. However, due to the linear dependence on number of pixels, this technique may still be out of bounds for near-term quantum computers because of the large size and detailed nature of images present in real-world use cases.

To address the above issues, in this work, we introduce an approach for dense approximate amplitude encoding of images as quantum states using a number of gates and qubits that are {\it logarithmic} in the number of pixels. This technique is based on converting a matrix product state representation of an image into a quantum circuit. A related application of this technique has been used for loading probability distributions to quantum states in \cite{iaconis2023quantum}. We note that tensor network methods have found applications in a wide range of quantum information problems and quantum machine learning in particular
\cite{huggins2019towards, stoudenmire2016supervised, novikov2017exponential, PhysRevX.12.011047, PRXQuantum.2.010342}, but not been used in the context of loading classical data to quantum states to our knowledge. Our method provides quantifiable control of the accuracy, which means that the fidelity of the encoded image can be systematically improved as quantum computers advance without changing the underlying encoding scheme.

We start by defining the amplitude encoding of images and how it can be used in a quantum machine learning algorithm. Suppose we have to encode a grayscale image with $N_x$ pixels on the $x$-axis and $N_y$ pixels on the $y$-axis. Let the image to be encoded be given by $(p_{xy}, x, y)$ with $1\leq x \leq N_x$ and $1\leq y \leq N_y$ and $0\leq p_{xy}\leq 1$ and $x, y \in \mathbb{N}$ are natural numbers. We can encode this image using $N=\log_2(N_x N_y)$ qubits defined as 
\begin{equation}\label{eq:image_state}
    \ket{\Psi}=\frac{1}{\mathcal{N}}\sum_{x=1}^{N_x}\sum_{y=1}^{N_y} \sqrt{p_{xy}}e^{i\phi_{xy}} \ket{x}\ket{y},
\end{equation}
where $\ket{x}$ and $\ket{y}$ are computational basis states corresponding to binary representations of $x$ and $y$ respectively, and $\mathcal{N}=\sum_{x=1}^{N_x}\sum_{y=1}^{N_y} p_{xy}$. $\phi_{xy}$ are arbitrary phases. For an image with multiple color channels, a state of this form can be used to encode each channel. 

A quantum machine learning model maps an input $(p_{xy}, x, y)$ to a value $c$ using a parametrized operator $\hat{C}$ which can then be used for further classification. If the model uses a single copy of $\ket{\Psi}$,

\begin{align}
    c=\sum_{x',x=1}^{N_x}\sum_{y',y=1}^{N_y} \sqrt{p_{x'y'}p_{xy}} e^{i(\phi_{xy}-\phi_{x'y'})}C_{x',y',x,y},
\end{align}
where $C_{x',y',x,y}=\bra{y'}\bra{x'}\hat{C}\ket{x}\ket{y}$. Thus, the output of the model is quadratic in the pixel amplitudes $\sqrt{p_{xy}}$. If $k$ copies of the state $\ket{\Psi}$ are used by the model, then the output is a polynomial of power $2k$ in the pixel intensities. Setting the total number of pixels, $L^2=N_x N_y$, we see that classically simulating this same model would thus involve a number of operations that scales as $O(L^{4k})$. Parametrized quantum circuits for machine learning have a number of gates that scale polynomially in the number of qubits, meaning that their execution time scales as $O(\log(L))$. With loading time linear in the number of pixels as in the QPIXL formalism, each execution of a quantum model would involve a number of operations that scales as $O(L^{2k})$, being limited by the loading time. The technique we present has loading time that is $O(\log(L))$ for a given accuracy and this allows for the execution time of a quantum model to also be $O(\log(L))$. Therefore, our loading technique allows potential quantum advantage from variational quantum models to clearly emerge, compared to previous loading methods where the execution time is dominated by the loading time.

We next set the stage for explaining our image loading technique by giving a brief introduction to matrix product states. A Matrix Product State (MPS) is a wave function of the form

\begin{eqnarray}
    \ket{\Psi} = \sum_{\{\sigma\}}\left[\prod_{i=1}^N M_{\alpha_{i-1} \alpha_{i}}^{[i],\sigma_i}\right] \ket{\sigma_1 \dots \sigma_N}
\end{eqnarray}
where the terms $M^{[i],\sigma_i}_{\alpha_{i-1},\alpha_i}$ are $N$ different 3-index tensors, and we use the Einstein summation convention that repeated indices are summed over. Each tensor contains a ``physical" index $\sigma_i \in [1,d]$, and ``bond" indices $\alpha_i \in [1,\chi]$ \cite{Schollwock_2011}. Here $d$ is the local dimension of the quantum state, so that $d=2$ for qubits. The maximum value of the bond indices $\alpha_i$ is known as the bond dimension $\chi$, and controls the amount of entanglement which can be represented by the MPS. A given MPS representation of a quantum state can be compressed by performing successive singular value decompositions (SVD) on the individual matrices and truncating the spectrum to eigenvalues/eigenvectors. For each matrix, the so called truncation error is given by the sum of the squares of the discarded singular values $\epsilon = \sum_{i=m+1}^{2^i}, \lambda_i^2$, and controls the fidelity of this compression method \cite{Schollwock_2011}. For a state $\ket{\psi}$ and allowed error $\epsilon$, we say that it can be approximately represented as a MPS if, for arbitrary $N$, there exists a fixed $\chi$ MPS,
$\ket{\tilde{\psi}}$ such that the Frobenius norm
\begin{eqnarray}
    ||\ket{\psi} - \ket{\tilde{\psi}}||^2 \le \epsilon.
\end{eqnarray}

In \cite{garcia2021quantum}, it was shown that smooth differentiable functions which are encoded in the amplitude of a quantum state using a big-endian binary encoding scheme have low entanglement, due to the vanishing additional entanglement cost of adding extra qubits of decreasing significance. This property was exploited in Ref's \cite{Holmes2020,iaconis2023quantum} to show that smooth 1D probability distributions can be efficiently loaded using MPS states, which leads to an efficient state preparation method for these distributions on a quantum computer.  While originally developed as efficient representations of 1D quantum states, matrix product states have since found rich applications when applied to 2D  and quasi-2D quantum systems \cite{stoudenmire2012studying}.  

\begin{figure*}
\centering
    
    \includegraphics[height=1.2in]{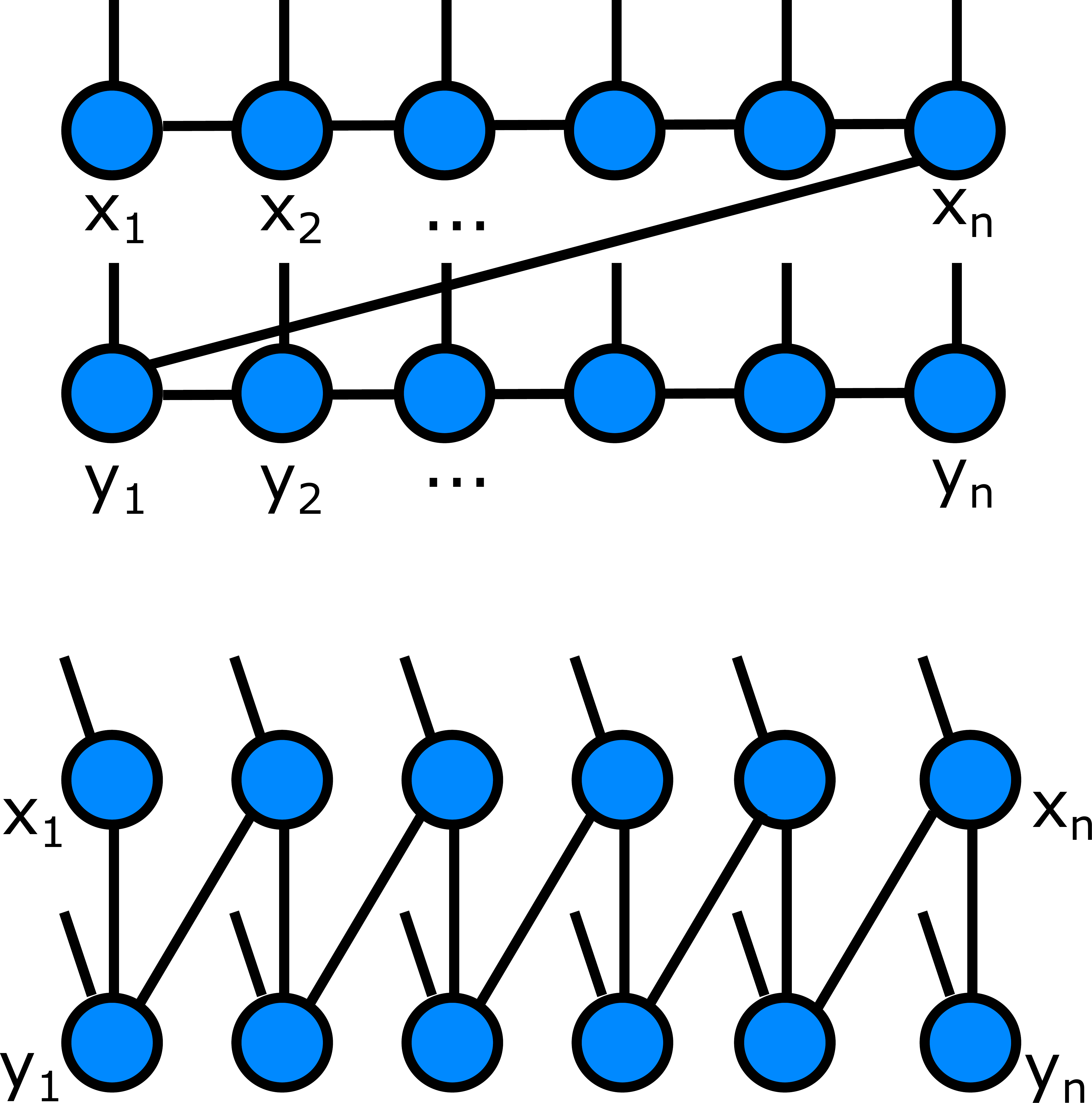}
    \quad \quad \includegraphics[height=1.2in]{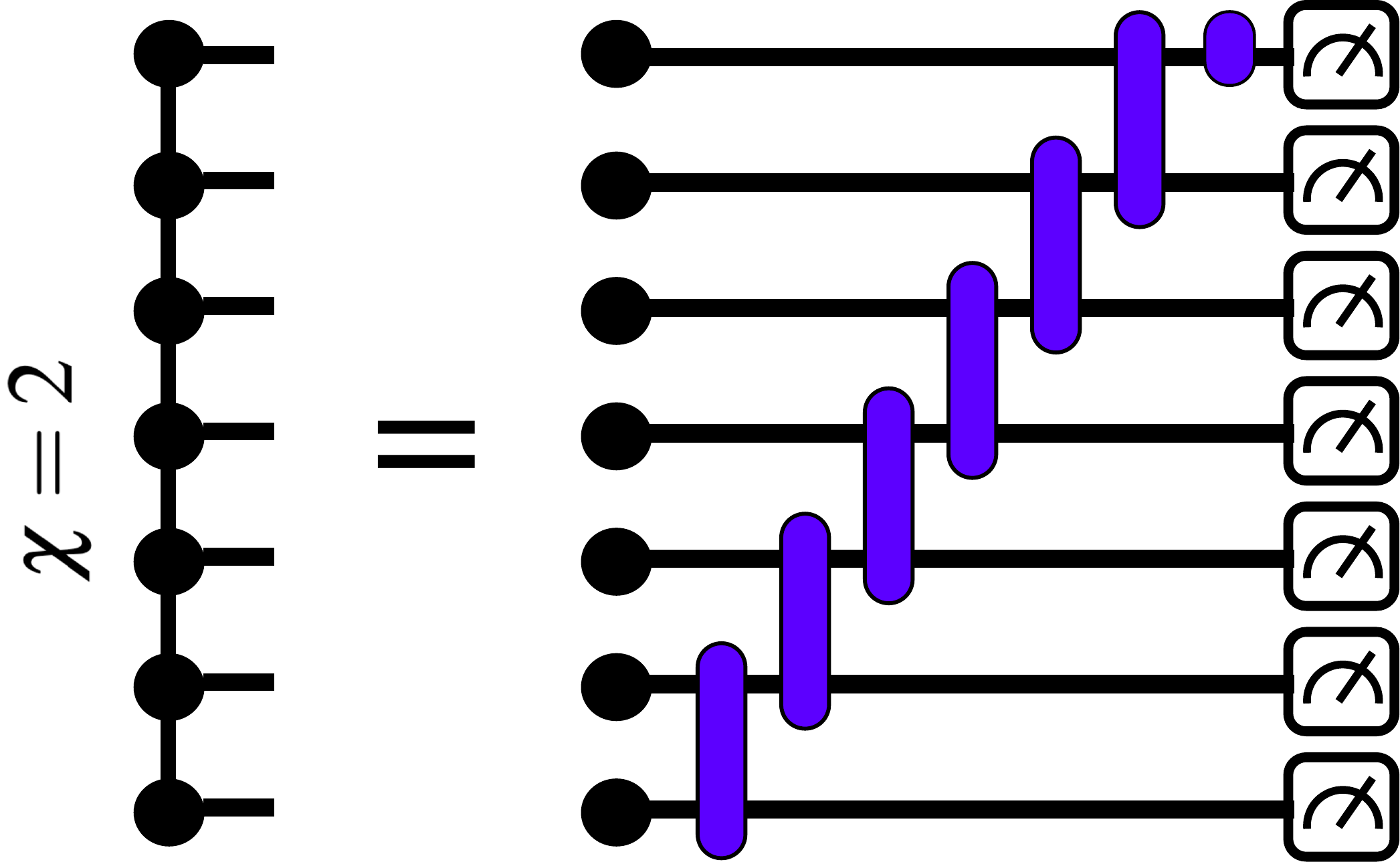}
    \quad \includegraphics[height=1.0in]{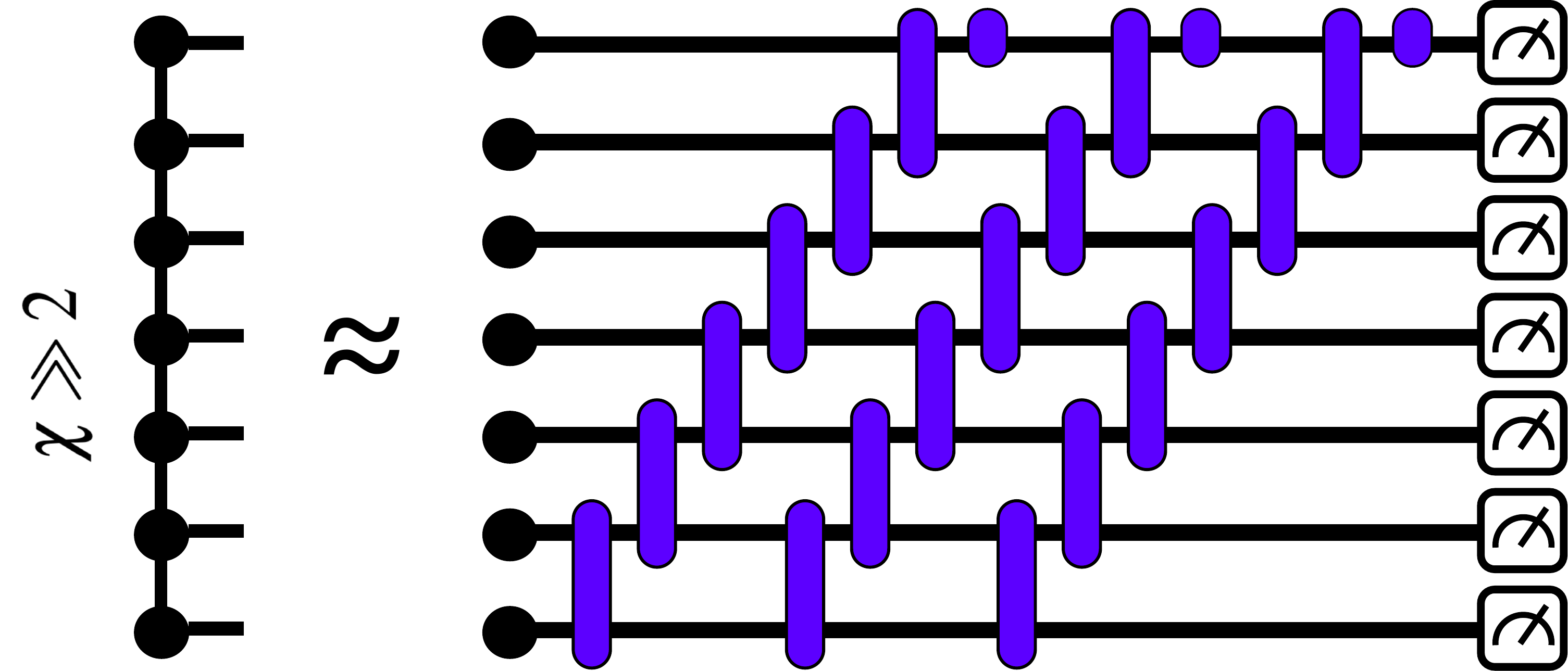}
    \caption{{(\it Left)} A 2D grayscale image whose pixel locations are represented by the binary $x$-$y$ pair $(x_1x_2\dots x_n,y_1y_2\dots y_n)$ can be encoded as a 2-leg ladder tensor network. A matrix product state representation can connect this ladder via one of the two 1D paths shown here. Throughout this work we show results from the upper figure, but find that both formulations give nearly identical results. {\it (Middle)} The quantum circuit representation of an MPS. At bond dimension 2, the MPS can be exactly represented by a single layer quantum circuit. {\it (Right)} For bond dimension $\chi>2$, the MPS can approximately be represented by $m$ layers of the $\chi=2$ circuit. }
    \label{fig:mps_circuit}
\end{figure*}

In this work, we demonstrate how these techniques also allow for an efficient representation of amplitude encoded 2D images by tensor networks, and therefore dramatically improve the prospects of encoding 2D images in quantum states. The amplitude encoded 2D image can be viewed as a quantum system on a 2-leg ladder, with the most significant digit of the $x$ and $y$ coordinate of the pixel location represented by the leftmost rung of the ladder, as in Fig.~\ref{fig:mps_circuit}. In the rest of the paper, we set $N_x=N_y=L$ but note that the construction is straightforwardly extended to the case $N_x\neq N_y$. 

We now describe our image loading procedure. State preparation of an arbitrary quantum state on $N$ qubits requires a circuit with $\mathcal{O}(2^N)$ CNOT gates.  A grayscale image with $L^2$ pixels can be efficiently represented using an amplitude encoded quantum state with only $N=2\log_2(L)$ qubits, however, exactly performing the state preparation procedure would require $L^2$ quantum gates. Instead, consider that if the indices $\alpha_{i-1} \in [1,m]$, $\alpha_i \in[1,n]$ and $\sigma \in [1,d]$, then  $M_{\alpha_{i-1},\alpha_i}^{\sigma_i}$ is an $m\times n\times d$ tensor which can be cast as an isometry from $m$ dimensions to $dn$ dimensions, with the property that
\begin{eqnarray}
    \sum_{\sigma,\alpha_{i}} M_{\alpha_{i-1},\alpha_i}^{\sigma_i} M^{* \sigma_i}_{\alpha_{i-1}'\alpha_i} = I_{\alpha_{i-1},\alpha_{i-1}'}.
\end{eqnarray}

Each of these isometries, $M_{\alpha_{i-1},\alpha_i}^{\sigma_i}$, can be implemented in a quantum circuit as an operator which acts on $\log_2(dn)$ qubits when $n\ge m$, which can be further decomposed in $\mathcal{O}(dmn)$ CNOT gates \cite{PhysRevA.93.032318}. When applied in series, as in the example for $\chi=2$ shown in Fig.~\ref{fig:mps_circuit}, the quantum circuit implementing these isometries exactly prepares the matrix product state wave function.  Therefore, a MPS with bond dimension $\chi$ consists of $\log_2(L)$ isometries with $m=n=\chi$ and can therefore be exactly prepared with only $\mathcal{O}(d \chi^2 \log(L) )$ gates. 
For small values of $\chi$, this represents a large compression in the circuit required for quantum state preparation. 

\begin{figure*}
\centering
    \includegraphics[width=2.8in]{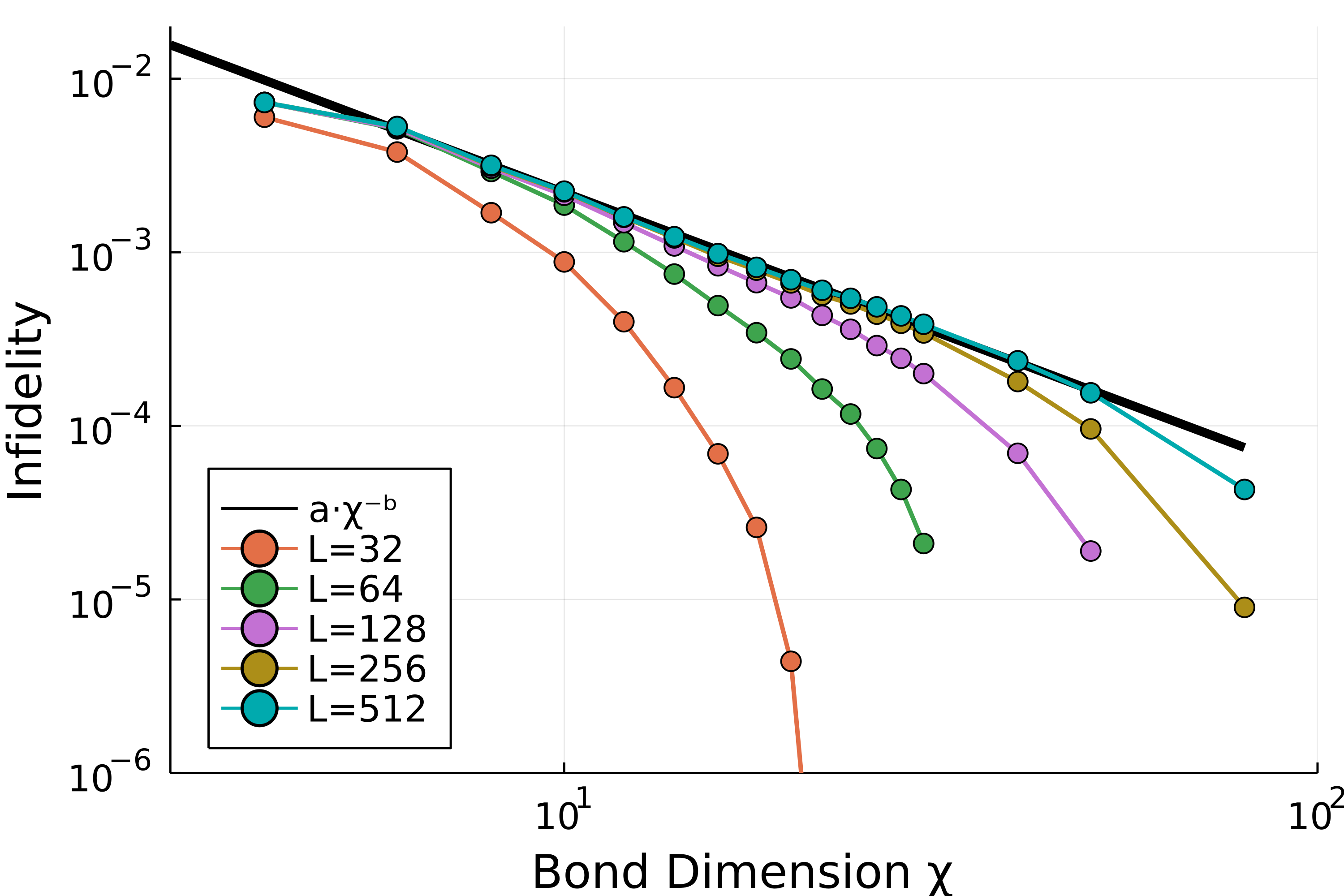}
    \quad \includegraphics[width=2.8in]{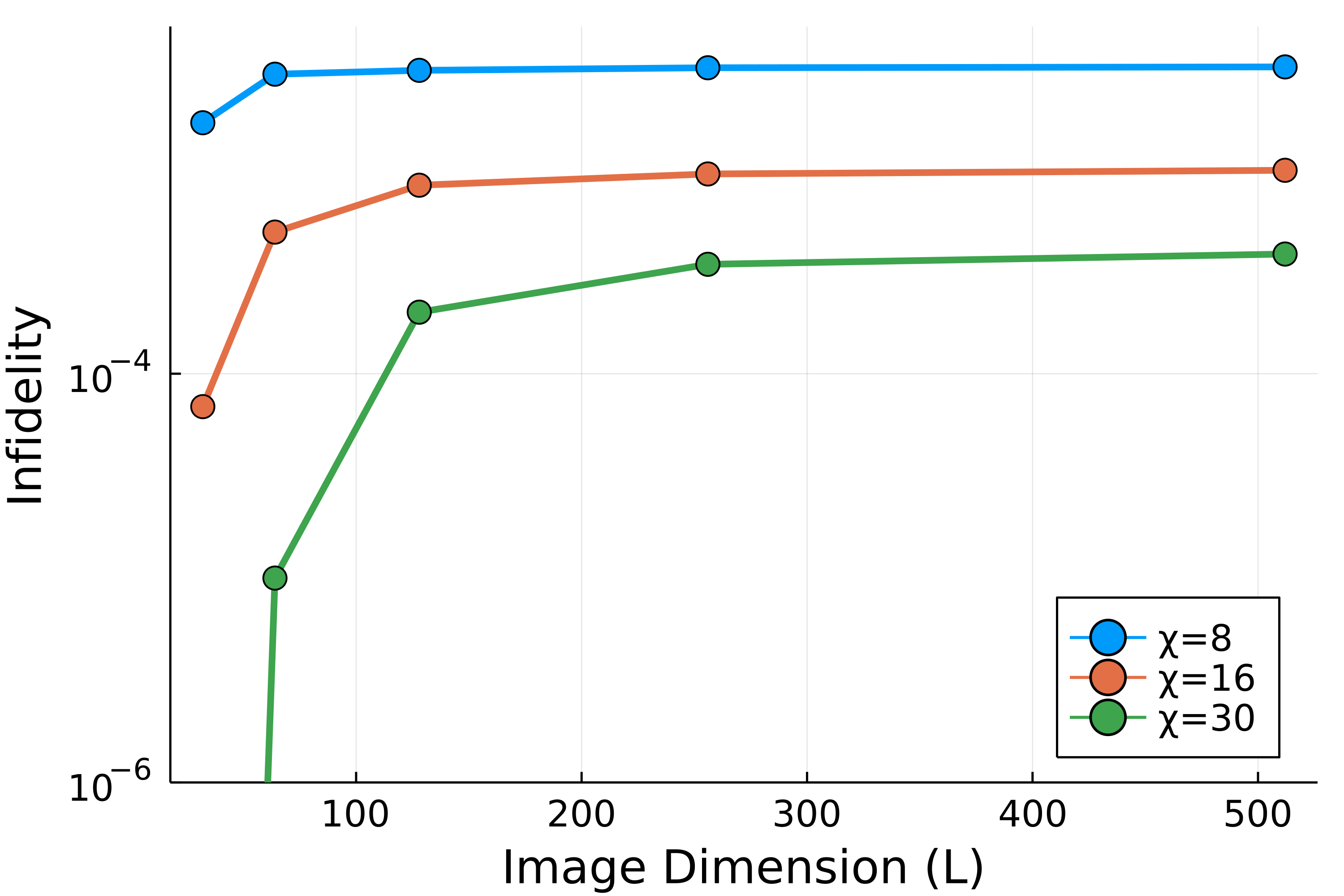}\\    \includegraphics[width=2.8in]{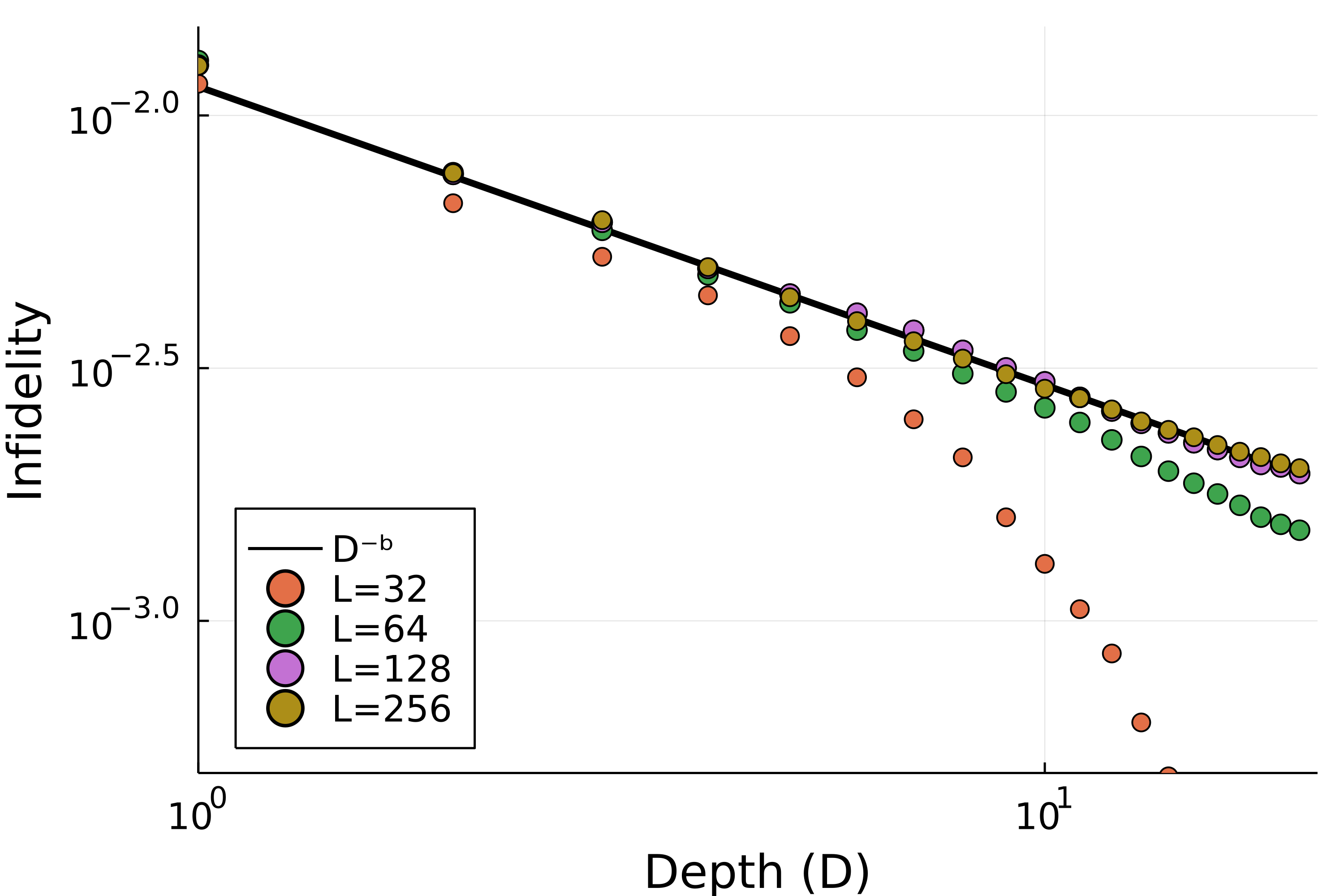}
    \quad \includegraphics[width=2.8in]{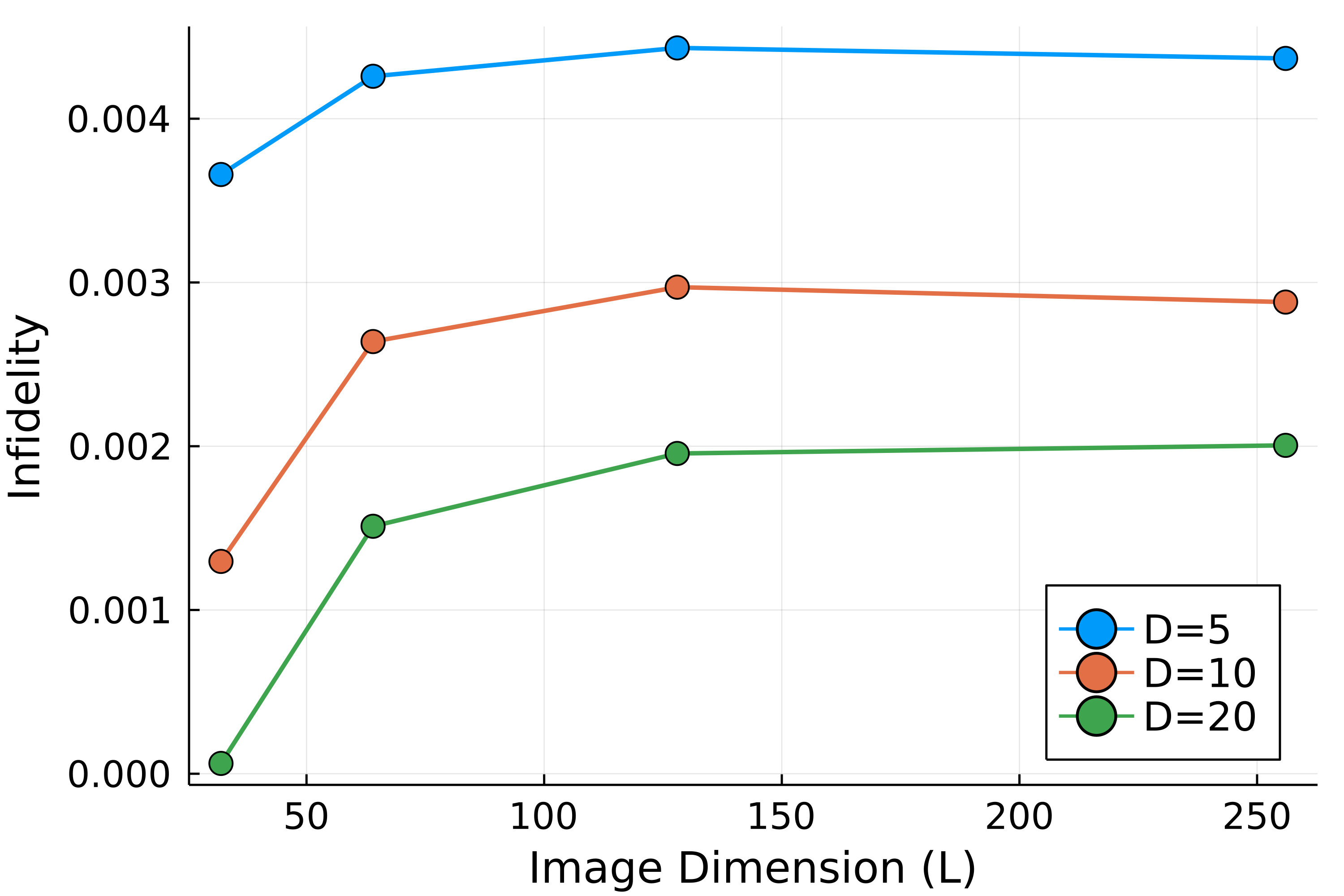}
    \caption{{\it (Top Left)} The infidelity of the tensor network approximation to the amplitude encoded image as a function of the bond dimension $\chi$ for copies of the same image at different resolutions of dimension $L\times L$. The infidelity appears to follow a power law when $\chi \ll L$ with $I \sim \chi^{-b}$. In this case we find that $b = 1.645 (18)$, although the exact value is likely dependent on the properties of the specific image. Note that the MPS approximation appears to perform better than expected on the smaller image size $L=32$. {\it (Top Right)} At fixed bond dimension, the overall infidelity appears  to saturate as a function of the image resolution $L$ as $L$ becomes large. This implies that high resolution images may be encoded to fixed accuracy using a number of gates which is only logarithmic in the image size. {\it (Bottom Left)} The infidelity of the quantum circuit reconstruction of the large $\chi$ state as a function of circuit depth for $L=32$ to $L=256$. Here we find that the infidelity again follows a power law as a function of circuit depth $D$ with $I \sim D^{-b}$ and $b=0.603(7)$. {\it (Bottom Right)}  At fixed circuit depth $D$, the infidelity saturates to a constant as a function of image dimension.}
    \label{fig:fidelity_scaling}
\end{figure*}

For images, we find that a constant $\chi$ is sufficient to represent an image to a given fidelity $|\langle \Psi|\tilde{\Psi}\rangle|$, independent of the image resolution. This is seen in the top right quadrant of Fig.~\ref{fig:fidelity_scaling}, where the infidelity $I=1-|\langle \Psi|\tilde{\Psi}\rangle|$ plateaus as a function of image size at fixed $\chi$. This implies that a circuit with a fixed depth and only $\mathcal{O}(\chi^2\log(L))$ gates can load arbitrarily high resolution images.  

\begin{figure*}
    \centering
    \includegraphics[width=6.0in]{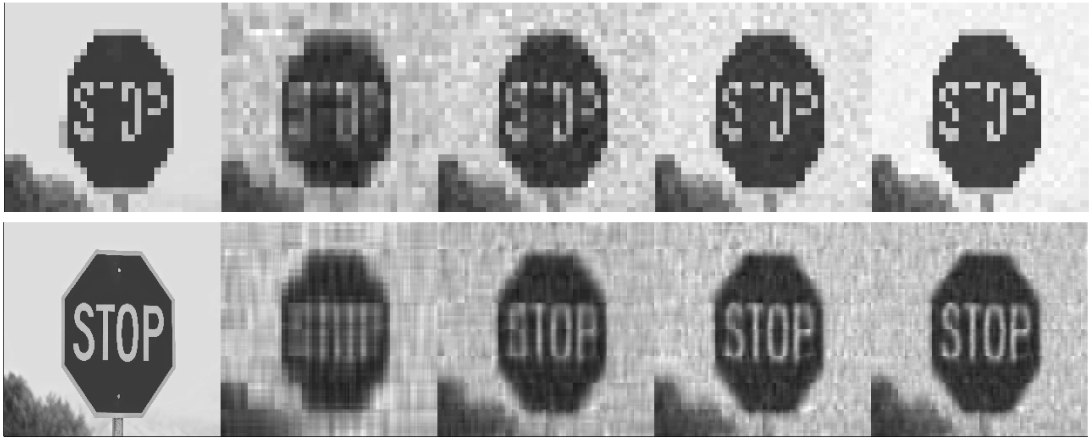}
    \caption{The exact {\it(Far Left)} and MPS reconstructed images of a stop sign for resolutions $L=32$ and $L=256$, using the iterative circuit construction at depths {\it (Left to Right)} $D=5,10,15$ and $20$ on an ideal simulator.}
    \label{fig:stopsign}
\end{figure*}

In Fig.~\ref{fig:fidelity_scaling} top left, we also demonstrate the infidelity  as a function of the truncated bond dimension $\chi$, when applied to images of a stop sign shown in Fig.~\ref{fig:stopsign}, which is down-scaled to a lower resolution of size $L\times L$. At large $L$, and for $\chi \ll L$, we find that the infidelity obeys the scaling law
\begin{eqnarray}
    I = \frac{a}{\chi^b}.
\end{eqnarray}
In this case, we find $b=1.645(18)$, although we expect that this exponent may depend on the specific properties of the image being encoded. Therefore, for high resolution images there is always a large compression which can be achieved using the MPS representation of the image, and the desired fidelity can always be increased by increasing the bond dimension $\chi$. Note, also, that as $\chi \rightarrow L$, the infidelity deceases more rapidly with $\chi$.

However, for near-term application, this MPS state preparation procedure still results in a large number of 2-qubit gates, which may render it impractical. For this reason, a number of approximation methods have been developed for directly constructing a high bond dimension MPS state using a small number of one and two qubit gates. A $\chi=2$ MPS can be simply prepared with a single layer quantum circuit of the form shown in Fig.~\ref{fig:mps_circuit}, where each of the $N$ two qubit unitaries applies a general $O(4)$ rotation that can be implemented using at most $2$ CNOT gates plus a potential SWAP operation \cite{vatan_williams}. For higher bond-dimension MPS states, there exist several low-depth state preparation algorithms which approximate the $\chi\gg 2$ MPS by repeatedly applying a series of single layer $\chi=2$ circuits.  The two main algorithms which have been proposed for this approximation are the iterative circuit construction \cite{ran2020encoding} and the gate-by-gate optimization method of Ref.\cite{shirakawa,zapata_mps}.   In the iterative circuit construction method, $D$ layers of $\chi=2$ MPS circuits are applied in series, with each layer $i$ implementing the unitary circuit $U^{[i]}$ such that
\begin{eqnarray}
    U^{[1]}\ket{0} &=& \ket{\Psi}\big|_{\chi=2} \label{eq:disentangle1}\\
    U^{[i]}\ket{0} &=& \left(U^{[i-1]\dagger}\dots U^{[1] \dagger} \ket{\Psi}\right)\bigg|_{\chi=2}\label{eq:disentangle2}\\
    U_{tot} &=& U_D \dots U_2 U_1 \label{eq:disentangle3}
\end{eqnarray}
where $\ket{\Psi}|_{\chi}$ represents the truncation of the wave function $\ket{\Psi}$, to bond dimension $\chi$. In this way, a MPS with high bond dimension $\chi$ can be approximately prepared using the circuit structure shown in Fig.~\ref{fig:mps_circuit}.  
In Ref.\cite{zapata_mps}, the same circuit structure is used to represent the quantum state, however it was shown that the individual unitaries can be optimized one at a time by sweeping through the quantum circuit and applying the optimization algorithm of Ref.~\cite{shirakawa}.  
In this optimization scheme, as explained in Ref.\cite{zapata_mps}, one calculates the environment tensor in a circuit with $M$ two-qubit unitary gates $U_i$ as
\begin{eqnarray}
    F_m = Tr_{\bar{U}_m} \left[\prod_{i=M}^{m+1} U_i \ket{\Psi} \bra{0} \prod_{j=1}^{m-1}U_j^\dagger \right],
\end{eqnarray}
where $Tr_{\bar{U}_m}$ is the trace over all qubit indices which don't interact with $U_m$ and which can be evaluated in practice by contracting the quantum circuit with gate $m$ removed with the exact MPS state $\ket{\Psi}$. The optimization algorithm proceeds by performing the SVD $F_m = \mathcal{U}\mathcal{S}\mathcal{V}^\dagger$, and replacing the unitary $U_i$ with the new unitary matrix $\mathcal{U}\mathcal{V}^\dagger$.
This sweeping gate-by-gate optimization algorithm can lead to a large improvement in the fidelity of the reconstructed state at the same circuit depth as the iterative algorithm.  In the rest of this work, we apply the improved version of this algorithm described in Ref.~\cite{zapata_mps}, where all $D-1$ layers of the MPS circuit are optimized using this gate-by-gate method, and a new layer numbered $D$ is generated using Eqs \ref{eq:disentangle1}-\ref{eq:disentangle3}. Once this new layer is added, all $D$ layers are re-optimized using the algorithm to produce the final circuit.

We use this circuit approximation method to generate low-depth quantum circuits which approximately prepare the quantum states in Eq. \ref{eq:image_state}. Throughout the optimization procedure, the bond dimension of the target MPS is limited to $\chi \le 32$. The results of this procedure as a function of circuit depth and image resolution are shown in Fig.~\ref{fig:fidelity_scaling}, for the sample stop sign image. Again, we find that the infidelity decreases as a power law with the circuit depth $D$, measured in terms of the number of MPS layers applied. We find that 
\begin{eqnarray}
    I = \frac{a}{D^b}
\end{eqnarray}
with $b=0.603(7)$. We also find that the infidelity tends to a constant as $L$ increases at fixed $D$.

In Fig.~\ref{fig:stopsign}, we show the reconstructed image which is generated by the quantum circuit at depths $D=5-20$. Again, we find an additional boost in performance when applied to images with the smallest resolution, so that for images of size $32\times32$, the reconstruction appears adequate at $D=10$ when $180$ CNOT gates are applied and nearly ideal at depth $20$, when we apply $360$ CNOT gates. This is a compression of roughly $80-90\%$ compared to a naive encoding method. We also show the reconstruction for images of dimension $256\times256$. Again, we see a good reconstruction between depths 10-20, where we apply 260-520 CNOT gates to load an image with 65536 pixels, a compression of $>99.5\%$.  

From Fig.~\ref{fig:stopsign}, we also see that the effect of increasing the circuit depth is to increase the sharpness of the edges. A similar effect is expected when the bond dimension $\chi$ is increased. This indicates that the use of larger depth circuits for encoding will increase the accuracy of visual pattern recognition and detection of small objects.

\begin{figure*}
    \includegraphics[width=3.0in]{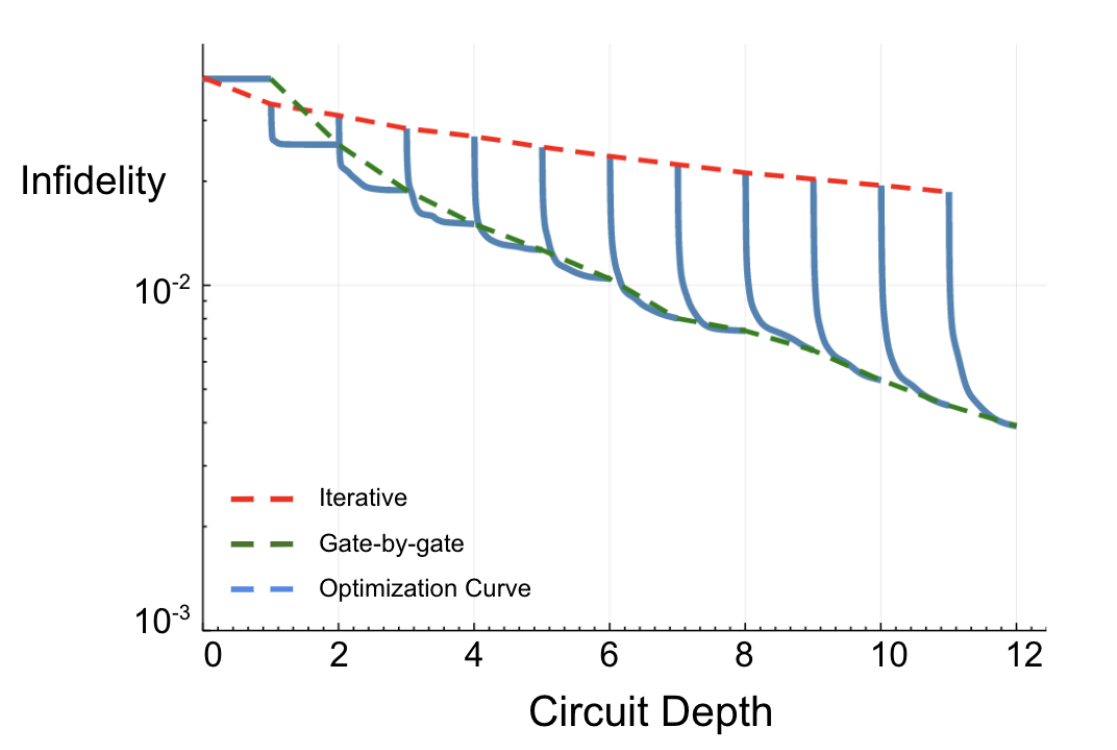}
    \includegraphics[width=3.6in]{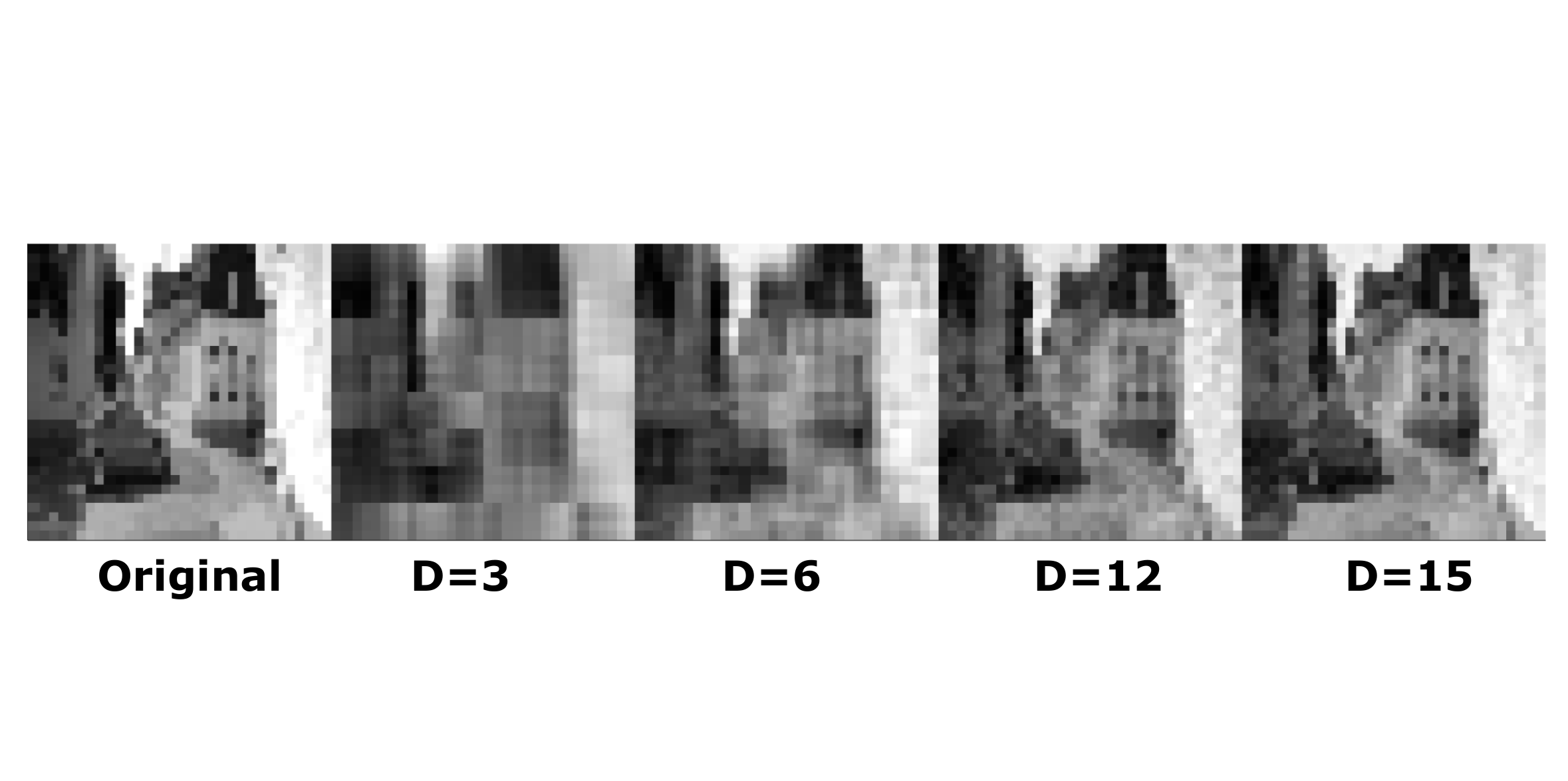}\\
    \caption{{\it (Left)} The scaling of the infidelity between the exact image and the MPS circuit approximation using two different circuit construction methods. The blue curve represents the infidelity throughout 200 sweeps of the gate-by-gate optimization algorithm at each circuit depth, where the initial circuit (and therefore initial infidelity) for each depth is the output of the iterative circuit construction algorithm. We see that the gate-by-gate optimization method proposed in Refs \cite{zapata_mps,shirakawa} significantly improves the fidelity compared to  the iterative loading method.  {\it (Right)} Reconstruction of the quantum state generated at different circuit depths on an ideal simulator.}
    \label{fig:mps_training}
\end{figure*}

In Fig.~\ref{fig:mps_training}, we perform the state preparation procedure on a more complex road scene image of size $32\times 32$. Even using this complex image, we see that our state preparation method gives a good reconstruction for depths $D>12$. We compare the iterative and gate-by-gate training procedures, where each gate-by-gate optimization is run for 200 sweeps. The starting point for the gate-by-gate optimization at each depth is the iterative circuit construction. We see that the gate-by-gate optimization gives a large improvement over the iterative circuit construction and appears to give superior scaling of Infidelity with circuit depth. 

\begin{figure*}
    \includegraphics[width=6.5in]{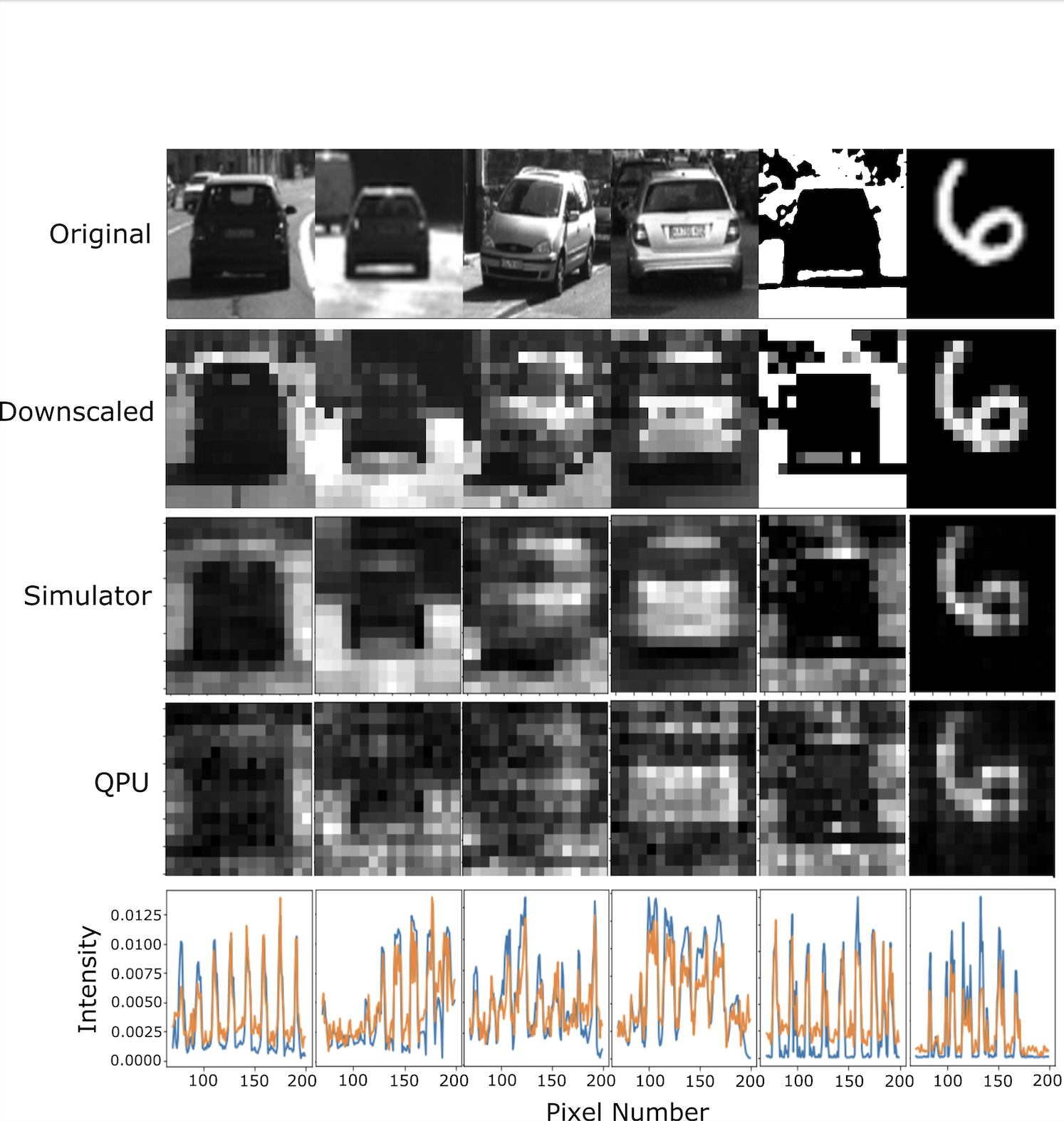}
    \caption{Demonstration of our image data loading algorithm on a trapped Ion Quantum computer, with images of resolution $16\times 16$. {\it (Top Row)} The original image at resolution $256\times256$. {(\it Second Row) } The original image downscaled to size $16\times16$. {\it (Third Row)} The output histogram of the MPS quantum circuit measured by the ideal simulator on 8 qubits, reshaped into a 2D image. {\it (Fourth Row)} The output histogram of the MPS quantum circuit measured by the trapped ion quantum computer on 8 qubits. {\it (Bottom Row)} The 1D wave function amplitudes on the noiseless simulator (blue curve) and the QPU (orange curve). Each circuit consists of 3 MPS circuit layers, implying that the image is loaded using only 42 CNOT gates, a significant compression compared to the 256 CNOTS required using a naive encoding. For each circuit, both on the ideal simulator and the trapped ion QPU, 10000 shots were taken in the computational basis. }
    \label{fig:qpu_results}
\end{figure*}

Finally, we experimentally implement this state preparation procedure on IonQ Aria, a trapped-ion quantum processing unit (QPU). IonQ Aria has been reported to have 25 `Algorithmic Qubits', which is a measure of quantum computing performance based on executing a set of application-oriented benchmarks \cite{qedc}. We choose 4 images of road scenes containing a vehicle and one simpler image from the MNIST dataset. Each original image is first down-scaled to dimension $16\times16$. We prepare each amplitude encoded state using a depth 3 MPS circuit with the gate-by-gate optimization method. These are then loaded onto 8 qubits of IonQ Aria one at a time. After loading, the qubits are measured in the computational basis to reconstruct the loaded image. The results of the reconstruction are shown in Fig.~\ref{fig:qpu_results} and compared to the original image as well as the results from an ideal simulator. Here the top row shows the original images and the second row the downscaled ones. The third row is the output from running the circuits on an ideal simulator. The fourth row is the output from the QPU. Note that in addition to noise due to gate execution and measurement, the reconstruction is also limited by statistical shot noise due to the limited number of shots that were taken.

We see that our state preparation method is able to maintain the large scale structure of the complex road scene images despite the noise on the hardware. To see this more clearly, in the bottom row, we plot curves that show the intensity as a function of pixel number, effectively `flattening' the image. We see that the QPU output closely follows the trend of the simulator output despite the noisy execution. The results are even more visually impressive when applied to the simpler MNIST image shown in the far right column of Fig.~\ref{fig:qpu_results}. As far as we are aware, this represents the first example of full amplitude encoding of large uncompressed images into quantum states. In all cases, we believe that enough detail in the images is maintained to be directly relevant for current quantum image processing algorithms.

In conclusion, we have proposed and demonstrated for the first time a technique for encoding images into quantum states that makes efficient use of the qubits as well as scales logarithmically in the number of pixels. We have shown through numerical testing that the technique has favorable scaling properties in terms of the circuit depths required to reach desired fidelities. Being suitable for near-term quantum computers, the technique allows for the design and testing of quantum learning models that are based on amplitudes of computational basis states. Since the data loading time is logarithmic in the number of pixels, it is no longer the leading order factor in the execution time of a typical quantum learning algorithm, and it allows for the realization of quantum advantage based on parameterized quantum circuits with the appropriate expressivity.

We expect that future work will improve the optimization procedures that are involved in creating the circuit. The exploration of higher-dimensional tensor network methods is also expected to make improvements to loading image fidelity while keeping the circuit depth the same. Our method will likely also generalize to other data types, such as video or three-dimensional data. Finally, we anticipate that an end-to-end demonstration of a quantum machine learning algorithm that utilizes this data loading scheme will lead to a future milestone in the field of quantum machine learning. 

\section*{Acknowledgements}
The authors acknowledge support from Hyundai Motor Company.

\newpage

\balance
\bibliographystyle{unsrt}
\titleformat{\section}[display]
  {\normalfont\Large\bfseries}{\vspace{1.5em}\refname}{0pt}{\Large}
\afterpage{\clearpage \bibliography{reference}}
\end{document}